\documentclass[journal=apchd5,manuscript=article,layout=traditional]{achemso}

\usepackage{graphicx}
\usepackage{amsmath,amssymb,amsfonts}
\usepackage{booktabs}
\usepackage{tikz}
\usepackage[hidelinks]{hyperref}
\usetikzlibrary{arrows.meta,positioning,fit}
\newcommand{\doi}[1]{\href{https://doi.org/#1}{DOI: #1}}

\SectionNumbersOn

\title[Functional Depth by Repeated Phase Encoding]
{Deep Inverse-Designed Nanophotonic Processors with Structural Nonlinearity from Repeated Phase Encoding}

\author{Azka Maula Iskandar Muda}
\author{U\u{g}ur Te\u{g}in}
\email{utegin@ku.edu.tr}
\affiliation[Ko\c{c} University]
{Department of Electrical and Electronics Engineering, Ko\c{c} University, Istanbul, 34450, T\"{u}rkiye}


\begin{document}

\begin{abstract}
Passive nanophotonic scattering regions implement linear optical transformations, and cascading input-independent regions alone does not create functional depth because the resulting transformations collapse into a single linear operator. Here, we introduce repeated phase encoding between inverse-designed passive transformations to generate an input-conditioned multilayer optical map without interlayer photodetection. Each re-encoding introduces additional phase-dependent optical pathways, producing structural nonlinearity with respect to the encoded variables while every scattering region remains passive and linear in the optical field. Under a controlled MNIST depth sweep, classification accuracy increases from 83.53\% with one layer to 93.61\% with seven layers, whereas the input-independent passive control saturates at 86.34\%. The depth trend also persists in a time-multiplexed CIFAR-10 patch model. We further realize three jointly trained $16\times16$ transformations, each independently implemented as an inverse-designed nanophotonic region, with relative complex transmission errors of 7.63\%, 7.59\%, and 8.80\%. The reconstructed electromagnetic stack reaches 91.79\% accuracy after phase calibration, compared with 91.95\% for its surrogate model. These results establish repeated input encoding as a route to functional depth in compact inverse-designed nanophotonic processors.
\end{abstract}

\begin{center}
\includegraphics[width=3.2in]{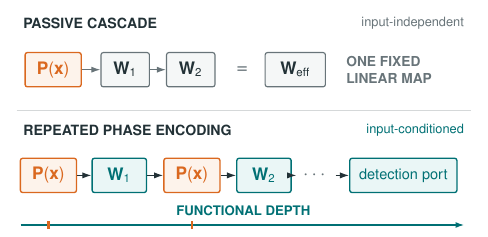}
\end{center}

\section{Introduction}

Photonic processors use propagation and interference to perform linear transformations in the optical domain. Programmable interferometer meshes, diffractive systems, wavelength-multiplexed accelerators, and integrated tensor cores support neural-network inference and convolutional workloads.\cite{shen2017,lin2018,xu2021,feldmann2021,ashtiani2022} Gradient-based learning has also been implemented on photonic hardware.\cite{pai2023,ashtiani2026} Inverse design can compress a trainable field transformation into a compact nanophotonic scattering region, but reported passive classifiers of this type have generally used one scattering stage.\cite{qu2020,sved2026,muda2026}

Cascading input-independent passive regions does not by itself create a deep input-output map. For transfer operators $\mathbf{W}_{\ell}$,
\begin{equation}
\mathbf{W}_{D}\mathbf{W}_{D-1}\cdots\mathbf{W}_{1}=\mathbf{W}_{\mathrm{eff}},
\label{eq:linear-collapse}
\end{equation}
so physical depth changes footprint and loss but the cascade remains one fixed linear operator. Interlayer detection and modulation can break this collapse,\cite{williamson2020,fard2020,zhou2022} but the optical state is then measured and a new field is prepared at every activation.

Here, \emph{structural nonlinearity} denotes nonlinear dependence of the end-to-end input-output map on the encoded data, rather than optical-material nonlinearity. Each passive scattering region remains linear in the incident optical field. We repeatedly apply an input-dependent unitary phase operator $\mathbf{P}(\mathbf{x})$ between passive transformations,
\begin{equation}
\mathbf{W}_{D}\mathbf{P}(\mathbf{x})\cdots\mathbf{W}_{1}\mathbf{P}(\mathbf{x}).
\label{eq:structural-depth}
\end{equation}
Because $\mathbf{P}(\mathbf{x})$ appears throughout the product, the transformation cannot be reduced to an input-independent $\mathbf{W}_{\mathrm{eff}}$ acting on a single encoding. The resulting system is linear with respect to the optical field, nonlinear with respect to the encoded variables, and functionally deep through repeated input conditioning.

Repeated data encoding has previously generated nonlinear input dependence in free-space linear optical systems.\cite{yildirim2024} Related data re-uploading schemes increase the expressivity of parameterized quantum circuits.\cite{perezsalinas2020} The contribution here is to establish repeated encoding in an integrated inverse-designed architecture, where the propagation stages are jointly trained as subunitary multiport transformations and selected transformations are subsequently realized independently as compact electromagnetic structures. We therefore ask not only whether repeated encoding increases model expressivity, but whether this depth survives the approximation, loss, reflection, and phase errors introduced by nanophotonic realization.

We answer this question in three steps. First, a controlled depth sweep tests whether additional passive transformations become functionally useful when the original input is re-encoded between them. Second, three selected $16\times16$ transformations are realized as inverse-designed nanophotonic regions. Third, the full-wave transfer matrices are reassembled into an end-to-end stack to quantify performance recovery, readout margins, and realization-error tolerance. The image-classification tasks serve as controlled probes of the physical architecture rather than as claims of state-of-the-art machine-learning performance.

\section{Results}

\subsection{Functional Depth through Repeated Phase Encoding}

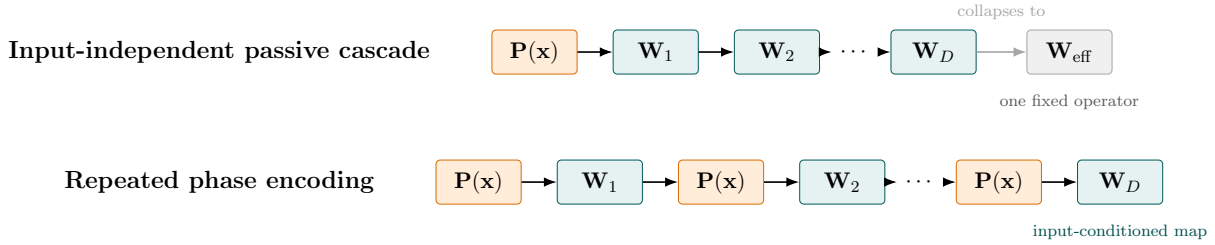
\begin{figure}[htbp]
\centering
\resizebox{0.98\linewidth}{!}{%
\begin{tikzpicture}[
  font=\sffamily\small,
  block/.style={draw,rounded corners=2pt,minimum width=1.45cm,minimum height=0.75cm,align=center},
  passive/.style={block,draw=teal!70!black,fill=teal!10},
  phase/.style={block,draw=orange!85!black,fill=orange!15},
  collapse/.style={block,draw=gray!70,fill=gray!12},
  arrow/.style={-{Latex[length=2.2mm]},thick},
  node distance=0.65cm and 0.6cm
]
\node[font=\bfseries] (l1) {Input-independent passive cascade};
\node[phase,right=0.9cm of l1] (p0) {$\mathbf{P}(\mathbf{x})$};
\node[passive,right=of p0] (w1) {$\mathbf{W}_{1}$};
\node[passive,right=of w1] (w2) {$\mathbf{W}_{2}$};
\node[right=0.20cm of w2] (dots1) {$\cdots$};
\node[passive,right=0.20cm of dots1] (wd) {$\mathbf{W}_{D}$};
\node[collapse,right=0.85cm of wd] (weff) {$\mathbf{W}_{\mathrm{eff}}$};
\draw[arrow] (p0)--(w1); \draw[arrow] (w1)--(w2); \draw[arrow] (w2)--(dots1); \draw[arrow] (dots1)--(wd);
\draw[arrow,gray!70] (wd)--node[midway,above=0.40cm,font=\scriptsize]{collapses to}(weff);
\node[below=0.2cm of weff,font=\scriptsize,gray!70!black] {one fixed operator};

\node[font=\bfseries,below=1.55cm of l1] (l2) {Repeated phase encoding};
\node[phase,right=0.9cm of l2] (q0) {$\mathbf{P}(\mathbf{x})$};
\node[passive,right=of q0] (u1) {$\mathbf{W}_{1}$};
\node[phase,right=of u1] (q1) {$\mathbf{P}(\mathbf{x})$};
\node[passive,right=of q1] (u2) {$\mathbf{W}_{2}$};
\node[right=0.20cm of u2] (dots2) {$\cdots$};
\node[phase,right=0.20cm of dots2] (qd) {$\mathbf{P}(\mathbf{x})$};
\node[passive,right=of qd] (ud) {$\mathbf{W}_{D}$};
\draw[arrow] (q0)--(u1); \draw[arrow] (u1)--(q1); \draw[arrow] (q1)--(u2); \draw[arrow] (u2)--(dots2); \draw[arrow] (dots2)--(qd); \draw[arrow] (qd)--(ud);
\node[below=0.2cm of ud,font=\scriptsize,text=teal!60!black] {input-conditioned map};
\end{tikzpicture}
}
\caption{Functional depth from repeated phase encoding. An input-independent passive cascade reduces to one fixed operator. Reintroducing the encoded phase between passive transformations makes the end-to-end operator input conditioned, even though every scattering region remains linear in the optical field.}
\label{fig:functional-depth}
\end{figure}

Figure~\ref{fig:functional-depth} contrasts an ordinary passive cascade with repeated input conditioning. In the conventional case, all transformations after the first encoding compose into the fixed operator in eq~\ref{eq:linear-collapse}. In the repeated-encoding architecture, the same data-dependent phase is reintroduced before every independently trainable passive transformation.

The nonlinear data dependence is explicit after only two layers. Taking the phase operator to be diagonal with entries $\exp(i\alpha x_j)$, the field at output channel $m$ is
\begin{equation}
a^{(2)}_m=\sum_{j,k}W_{2,mj}W_{1,jk}a^{(0)}_k
\exp\!\left[i\alpha(x_j+x_k)\right].
\label{eq:path-expansion}
\end{equation}
At depth $D$, each optical path accumulates a product of $D$ phase factors and thus a phase set by a sum of $D$ selected input coordinates. Interference among these paths generates progressively richer functions of the encoded variables. This mechanism requires no intensity-dependent material response: propagation remains linear for a fixed $\mathbf{x}$.

\subsection{Controlled Depth Scaling}

MNIST provides the cleanest controlled test of whether repeated encoding makes additional passive transformations useful. Sixteen principal-component features were mapped to optical phases, the width was fixed at 16 channels, and depth was varied while data, optical constraints, optimizer settings, seed, and checkpoint rule were held fixed across model families.

\begin{figure}[htbp]
\centering
\begin{minipage}{0.48\linewidth}
\centering
\includegraphics[width=0.90\linewidth,trim=56 446 1128 29,clip]{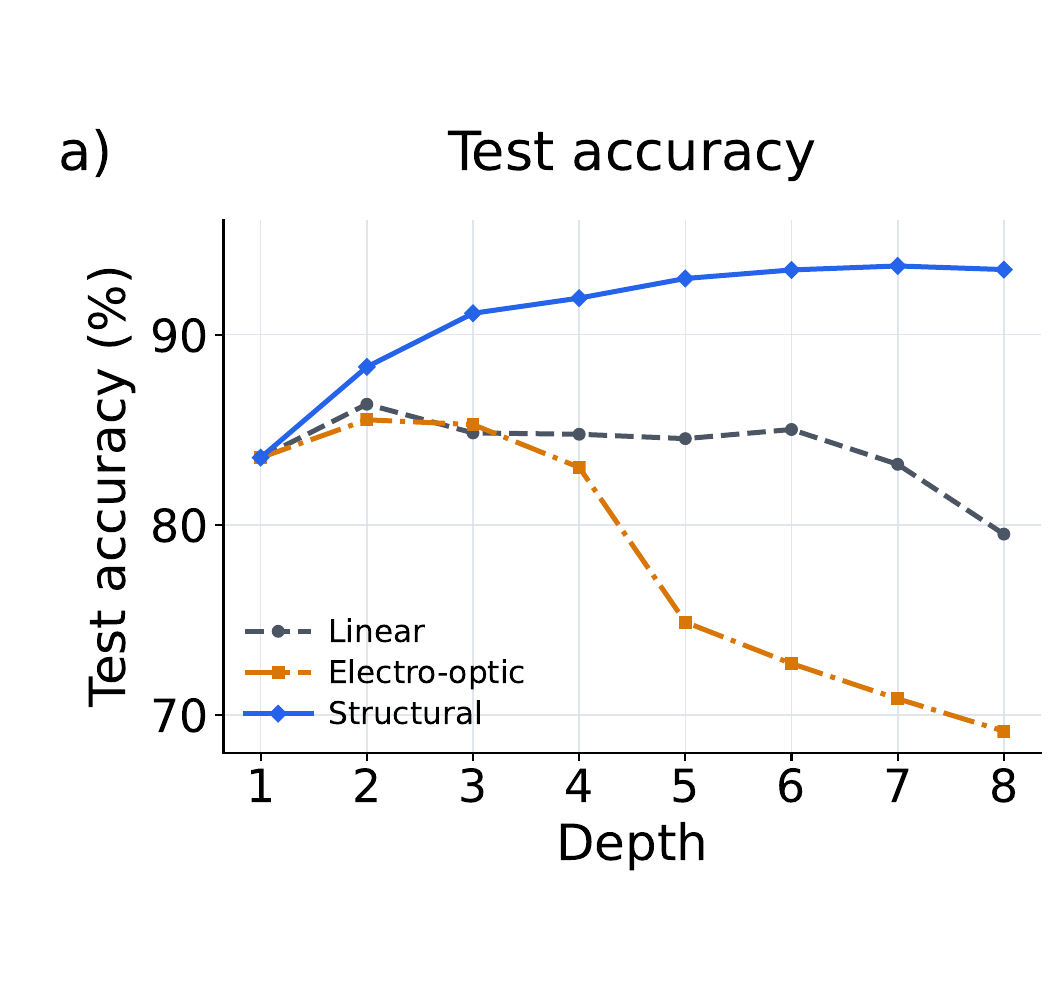}
\end{minipage}\hfill
\begin{minipage}{0.48\linewidth}
\centering
\includegraphics[width=0.90\linewidth,trim=56 0 1128 475,clip]{figure2.pdf}
\end{minipage}
\caption{Controlled depth scaling. (a) MNIST test-split accuracy under the common full-batch checkpoint protocol. (b) CIFAR-10 validation accuracy for the time-multiplexed patch model. Within each dataset, all three families share the data, optical constraints, optimizer settings, seed, and checkpoint rule. Loss and insertion-loss curves are provided in Supporting Information Figure~S2.}
\label{fig:depth-sweep}
\end{figure}

Under this common protocol, repeated phase encoding improved accuracy from 83.53\% at one layer to 91.12\% at three layers and 93.61\% at seven layers (Fig.~\ref{fig:depth-sweep}a). The input-independent linear control peaked at 86.34\% with two layers and then saturated or degraded. The essential result is therefore the relative depth trend under matched conditions: additional passive operators become functionally useful only when separated by repeated input conditioning.

The electro-optic model is included as an illustrative control, not as a comprehensive representation of electro-optic neural networks. Under the specified response, launch power, and passive-layer constraints, it peaked at 85.52\% and degraded at larger depth. At quadrature bias, its zero-power intensity transmission is $(1-t)/2=0.475$, equivalent to 3.23~dB per activation. As attenuation compounds, the intensity-dependent phase $g|a|^2$ also decreases and later stages approach a fixed complex scaling. Repeated encoding instead applies a unit-magnitude, data-dependent phase at every depth.

Because the standard MNIST test split was also used for checkpoint selection in this sweep, the absolute accuracies are not independently selected estimates of generalization. They should be interpreted as a controlled architectural comparison. This limitation does not change the matched depth trend, but it prevents treating 93.61\% as a stand-alone machine-learning benchmark.

\subsection{Inverse-Designed Nanophotonic Realization}

\begin{figure}[htbp]
\centering
\includegraphics[width=\linewidth]{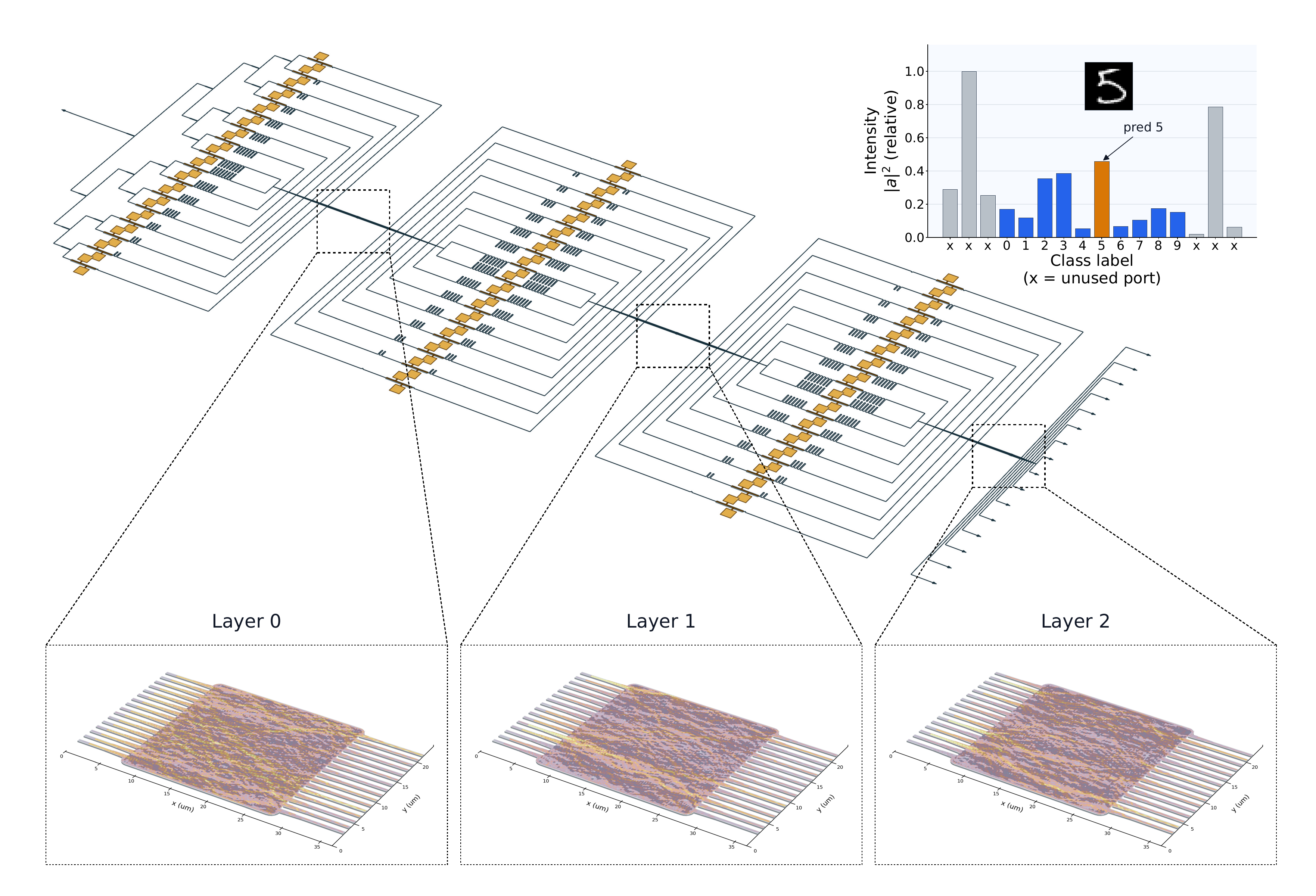}
\caption{Integrated three-layer realization and field propagation. The upper panels show the assembled circuit, including the repeated input-phase planes between three inverse-designed passive transformations, and the output for one MNIST example. The lower panels show the $21\times21~\mu\mathrm{m}^2$ design regions and propagated field magnitude for the three layers. The ten central output waveguides form the class readout.}
\label{fig:physical-stack}
\end{figure}

A separately trained three-layer MNIST model reached 91.95\% accuracy and supplied three complex target transformations for electromagnetic realization. Each transformation was implemented independently as a $21\times21~\mu\mathrm{m}^2$ inverse-designed region connecting 16 input and 16 output waveguides (Fig.~\ref{fig:physical-stack}). The repeated phase encoders are external to these passive multimode regions and reintroduce the same input before each stage.

\begin{figure}[htbp]
\centering
\includegraphics[width=\linewidth]{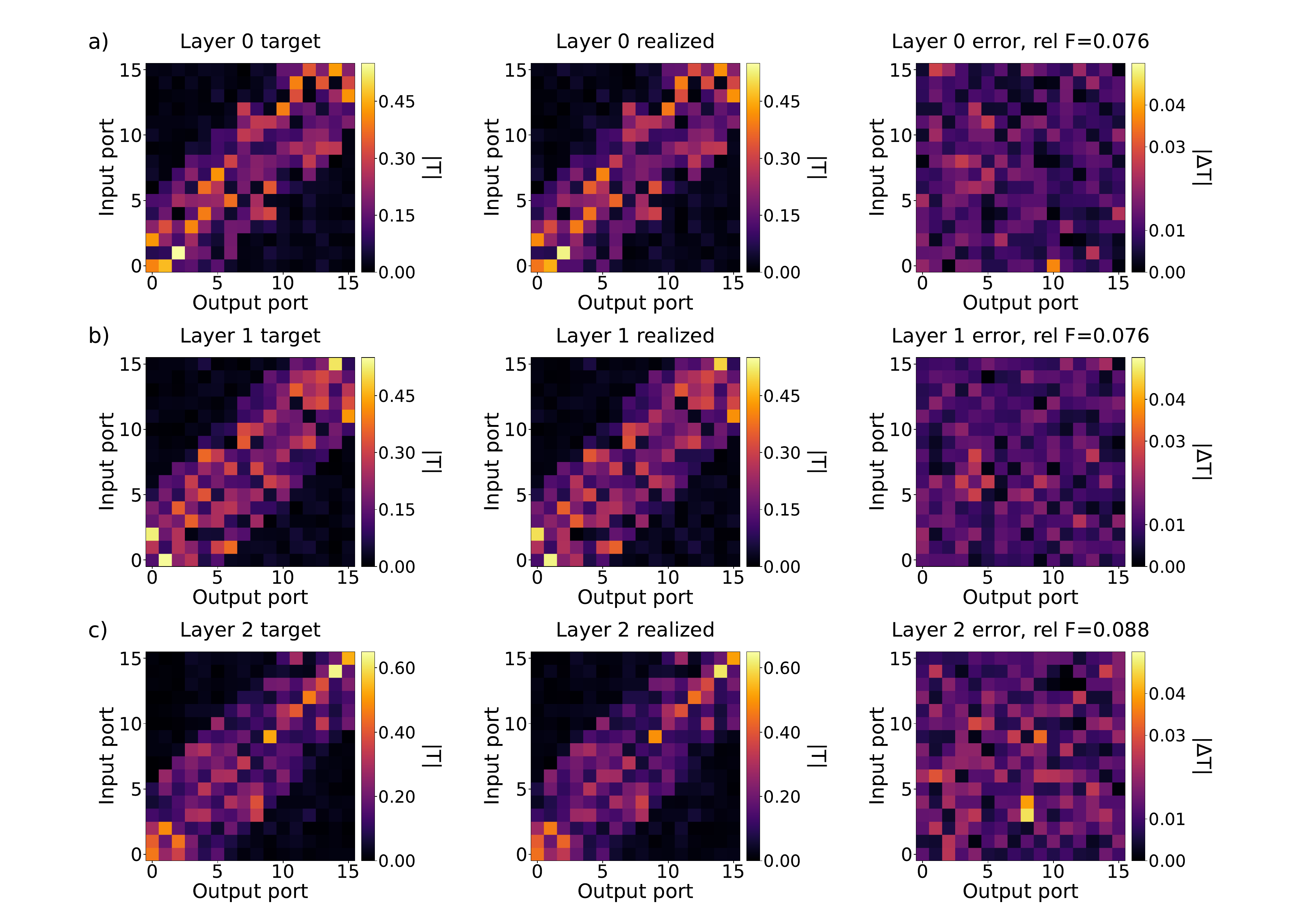}
\caption{Target-to-device transfer for the three inverse-designed regions. Columns show target magnitude $|T|$, realized magnitude $|T|$, and elementwise magnitude error $|\Delta T|$ for layers 0--2. Optimization and the stated relative Frobenius errors use the complex transmission matrices, including phase; only magnitudes are visualized here. Each $16\times16$ transformation is implemented in a $21\times21~\mu\mathrm{m}^2$ trainable region subject to the constraints specified in Methods.}
\label{fig:complex-transfer}
\end{figure}

The realized complex transmission matrices reproduce the dominant banded and local-coupling structure of their targets (Fig.~\ref{fig:complex-transfer}). Relative complex Frobenius errors were 7.63\%, 7.59\%, and 8.80\% for layers 0, 1, and 2. Their insertion losses were 4.36, 4.35, and 4.13~dB, within the 3--6~dB range used to construct the targets, while reflection errors were 4.83\%, 3.93\%, and 3.98\%. Although Figure~\ref{fig:complex-transfer} displays magnitudes for visual clarity, both inverse-design optimization and the reported errors use the full complex matrices.

\subsection{End-to-End Performance and Tolerance}

\begin{figure}[htbp]
\centering
\includegraphics[width=\linewidth]{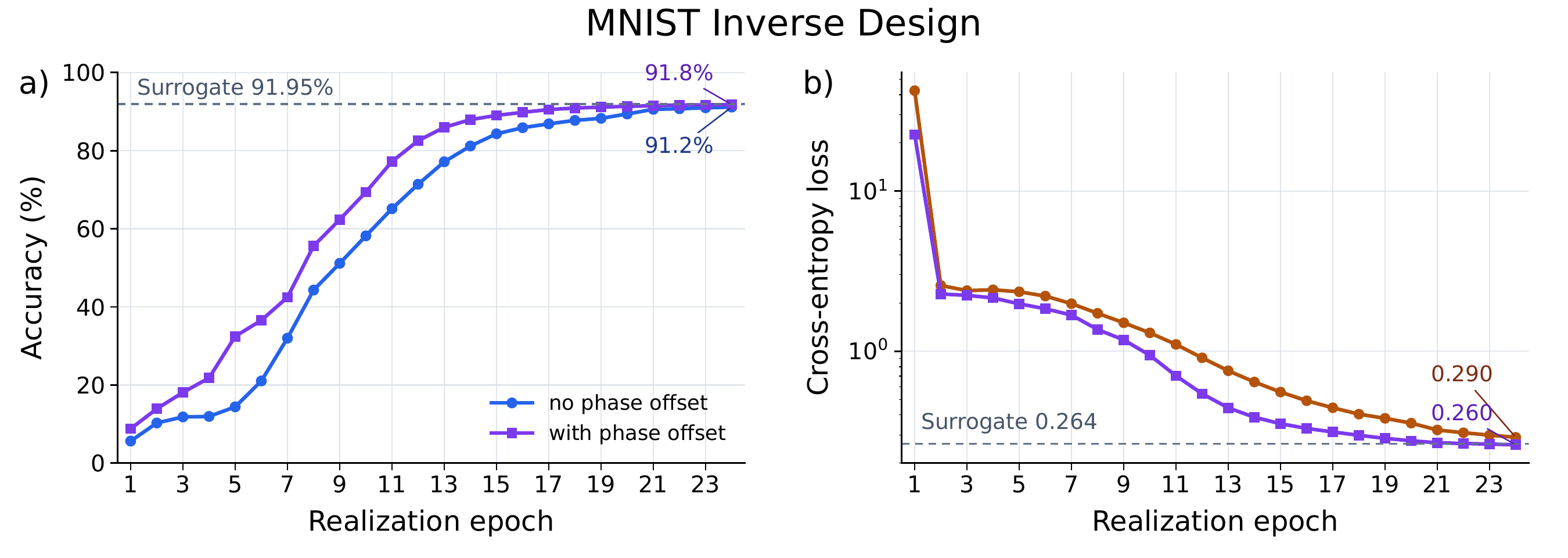}
\caption{End-to-end performance during electromagnetic realization. (a) MNIST accuracy of the reconstructed three-layer stack with and without layer-specific phase calibration. (b) Corresponding cross-entropy. Dashed lines mark the surrogate model. The final calibrated stack reaches 91.79\% accuracy, 0.16 percentage points below the surrogate.}
\label{fig:realization-recovery}
\end{figure}

The reconstructed electromagnetic stack reached 91.16\% accuracy before calibration and 91.79\% after optimizing layer-specific phase offsets on the training split, compared with 91.95\% for the surrogate model (Fig.~\ref{fig:realization-recovery}). The 0.63-percentage-point recovery indicates that coherent phase correction can compensate part of the aggregate transfer mismatch without modifying any passive region. Cross-entropy was 0.290 without phase offsets, 0.260 with offsets, and 0.264 for the surrogate.

\begin{figure}[htbp]
\centering
\includegraphics[width=\linewidth]{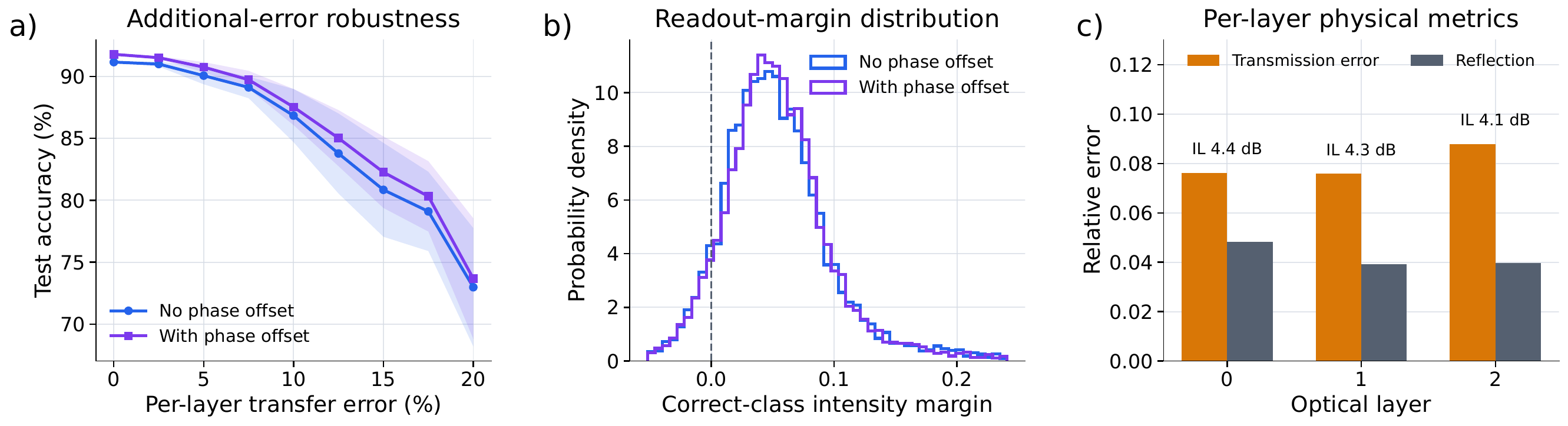}
\caption{Tolerance and physical metrics of the realized optical stack. (a) Accuracy after adding independent complex transfer perturbations to every layer; bands show one standard deviation across 20 seeds. (b) Correct-class intensity-margin distributions. (c) Relative complex transmission error, reflection error, and insertion loss of each inverse-designed layer.}
\label{fig:robustness}
\end{figure}

Independent complex perturbations produce gradual degradation before a sharper decline at larger transfer error (Fig.~\ref{fig:robustness}a). At 10\% added relative error per layer, accuracy was $86.84\pm2.14$\% without phase offsets and $87.53\pm1.46$\% with offsets across 20 perturbation seeds; at 20\%, the corresponding values were $72.98\pm4.78$\% and $73.68\pm4.87$\%. The calibrated stack also increased the mean correct-class intensity margin from 0.05374 to 0.05424 (Fig.~\ref{fig:robustness}b). The measured transmission error, reflection, and insertion loss of each layer are summarized in Fig.~\ref{fig:robustness}c. A single-sample readout and the full confusion matrix are moved to Supporting Information Figure~S3.

\subsection{Extension to Time-Multiplexed Image Patches}

To test whether the depth trend extends beyond compact principal-component inputs, we applied the same architecture to sequential $4\times4$ grayscale CIFAR-10 image patches. The optical outputs were spatially pooled and classified with an electronic ridge readout. This model is a generality check rather than a competitive CIFAR-10 classifier: it used 10,000 class-balanced training images, a 3,000-image validation subset, grayscale conversion, and an electronic linear readout.

The seven-layer repeated-encoding model reached 48.30\% validation accuracy, whereas the best linear and illustrative electro-optic checkpoints across the sweep reached 43.50\% and 45.00\% (Fig.~\ref{fig:depth-sweep}b). The result shows that the structural-depth trend persists under a different encoding and readout, but the restricted data protocol and electronic classifier preclude direct comparison with full-dataset CIFAR-10 benchmarks. Processing, validation-loss, and insertion-loss details are provided in Supporting Information.

\section{Discussion}

The central result is that repeated phase encoding prevents a cascade of passive nanophotonic transformations from collapsing into one input-independent linear operator. Functional depth arises because the original input conditions every layer, creating path-dependent phase products that interfere at the output. This mechanism is physically distinct from an intensity-dependent activation: the passive scattering regions remain linear in the optical field, and the nonlinear dependence is with respect to the encoded data.

The depth advantage also survives physical transfer from trained operators to inverse-designed electromagnetic structures. Three jointly trained $16\times16$ transformations were independently reproduced with 7.6--8.8\% relative complex error, and the calibrated full-wave stack retained 91.79\% accuracy relative to 91.95\% for the surrogate. The close agreement is more consequential for the architecture than the absolute classification score: it shows that the input-conditioned computation persists through approximation, insertion loss, reflection, and phase mismatch.

The architecture replaces interlayer state detection with repeated access to the original input. Its practical implementation therefore requires the encoded input to be broadcast or remodulated before successive passive transformations. The approach is most attractive in inference settings where the same input can be distributed across layers or where phase modulation can be time multiplexed. The passive character applies to the multimode transformations; the complete system includes external phase encoders. Accordingly, we describe the system as passive transformations separated by phase re-encoding and do not attribute passivity to the entire system.

Several limitations define the present evidence. The MNIST depth sweep uses the standard test split for checkpoint selection and reporting, so its absolute accuracy is not an independently selected test estimate. The CIFAR-10 study uses subsets, grayscale inputs, and an electronic ridge readout. The device study is based on two-dimensional effective-index simulations at one wavelength and polarization; it does not include three-dimensional modeling, fabrication variation, or measured hardware. These limitations leave fabrication and independent benchmarking as future tests, while the current calculations isolate the architectural mechanism and its survival under full-wave operator transfer.

\section{Conclusion}

Repeated input encoding provides a route to functional depth in integrated nanophotonic processors whose individual transformations remain passive and linear. By independently reproducing the jointly trained transformations as inverse-designed scattering regions and reconstructing their end-to-end response, we show that the resulting input-conditioned computation survives realistic transfer error, reflection, and insertion loss. This framework separates functional depth from optical-material nonlinearity and offers a compact architecture for multilayer nanophotonic inference without interlayer photodetection.

\section{Methods}

\subsection{MNIST Encoding and Controlled Depth Sweep}

We used the standard MNIST training and test splits.\cite{lecun1998} Each image was flattened and normalized to $[0,1]$. Principal-component analysis\cite{jolliffe2002} fitted on the 60,000-image training split retained 16 components, which were standardized with training-set statistics. The 10,000-image standard test split was used for both checkpoint selection and performance reporting in the controlled depth sweep. Components were assigned to physical ports in the order $(14,12,10,0,1,2,3,4,5,6,7,8,9,11,13,15)$.

All ports were launched with unit field amplitude, $\mathbf{a}_0=\mathbf{1}\in\mathbb{C}^{16}$. For standardized features $\mathbf{z}$, the input phase was
\begin{equation}
\mathbf{p}(\mathbf{z})=\exp(i\alpha\mathbf{z}), \qquad \alpha=0.5.
\label{eq:phase-encoding}
\end{equation}
The structural classifier propagated the field as
\begin{equation}
\mathbf{a}_{\ell}=\mathbf{W}_{\ell}\left[\mathbf{a}_{\ell-1}\odot\mathbf{p}(\mathbf{z})\right],
\qquad \ell\in\{1,\ldots,D\}.
\label{eq:optical-propagation}
\end{equation}
The linear control applied $\mathbf{p}(\mathbf{z})$ before the first layer only. The illustrative electro-optic control applied the response of Williamson \emph{et al.}\cite{williamson2020} between later passive layers,
\begin{equation}
\mathcal{F}(a)=\frac{i}{2}\sqrt{1-t}\left[1+\exp(-i\phi)\right]a,
\qquad \phi=g|a|^2+b,
\label{eq:williamson-response}
\end{equation}
with $g=0.05$, $b=\pi/2$, and $t=0.05$.

Each passive layer was parameterized as
\begin{equation}
\mathbf{W}_{\ell}=\mathbf{U}_{\ell}\operatorname{diag}(\boldsymbol{\sigma}_{\ell})\mathbf{V}_{\ell}^{\dagger},
\label{eq:subunitary-parameterization}
\end{equation}
where $\mathbf{U}_{\ell}$ and $\mathbf{V}_{\ell}$ are unitary and $10^{-6/20}\leq\sigma_{\ell,k}\leq10^{-3/20}$. This enforces passivity and 3--6~dB insertion loss per layer. A five-port routing window constrained transmitted power. For layer $\ell$, fractional out-of-window power was
\begin{equation}
\ell_{\mathrm{route},\ell}=\frac{1}{N}\sum_{j=1}^{N}
\frac{\sum_{i:\,|i-j|>5}|W_{\ell,ij}|^2}{\sum_{i=1}^{N}|W_{\ell,ij}|^2}.
\label{eq:routing-leakage}
\end{equation}
The depth-averaged leakage was $\ell_{\mathrm{route}}=D^{-1}\sum_{\ell}\ell_{\mathrm{route},\ell}$, and the surrogate objective was
\begin{equation}
\mathcal{L}_{\mathrm{sur}}=\mathcal{L}_{\mathrm{CE}}+5\max(0,\ell_{\mathrm{route}}-0.01)
+2\left[\max\left(0,\frac{\gamma-240}{240}\right)\right]^2.
\label{eq:surrogate-objective}
\end{equation}
Adam\cite{kingma2015} used 1,000 full-batch updates, a learning rate of 0.005, and seed 7 for every family and depth. The checkpoint with the highest test-split accuracy was retained at each depth. Because the objective was a matched architectural comparison rather than an absolute benchmark, every architecture used this same checkpoint protocol. Consequently, absolute accuracies should not be interpreted as independently selected test estimates; the relevant result is the relative depth trend under identical conditions. A separate three-layer structural run used 1,000 epochs with batch size 1,000 and supplied the electromagnetic targets.

The ten central output ports, indices 3--12, formed the optical readout: $I_c=|a_{D,c+3}|^2$. During training, a positive inverse-temperature $\gamma$ scaled the logits, $o_c=\gamma I_c$, but did not change the predicted class because $\arg\max_c\gamma I_c=\arg\max_c I_c$. It is absent from optical detection.

\subsection{Time-Multiplexed CIFAR-10 Patch Model}

RGB CIFAR-10 images\cite{krizhevsky2009} were converted to grayscale by $g=0.299R+0.587G+0.114B$. Overlapping $4\times4$ patches with stride two produced 225 vectors $\mathbf{x}_q\in\mathbb{R}^{16}$. Each patch was processed sequentially, with structural propagation
\begin{equation}
\mathbf{a}_{\ell,q}=\mathbf{W}_{\ell}\left[\mathbf{a}_{\ell-1,q}\odot
\exp(i\pi\mathbf{x}_q)\right],\qquad \mathbf{a}_{0,q}=0.25\mathbf{1}.
\label{eq:cifar-patch-propagation}
\end{equation}
Nonoverlapping $3\times3$ spatial averaging reduced the $15\times15$ patch lattice to $5\times5$ cells. The resulting 400 intensity values were standardized with training-set statistics and classified by ridge regression with coefficient $10^{-3}$. Training used 10,000 class-balanced training images and a separate 3,000-image validation subset. Adam ran for 80 epochs with batch size 64, learning rate 0.01, and seed 0. For each depth and family, the reported checkpoint maximized validation accuracy among epochs 1, 5, 10, 25, 50, 75, and 80. Full objective and readout equations are provided in Supporting Information.

\subsection{Electromagnetic Realization}

We followed a two-stage surrogate inverse-design workflow.\cite{muda2026} Task learning first produced passive complex targets. Each target was then realized independently in Tidy3D by an adjoint finite-difference time-domain workflow with automatic differentiation.\cite{piggott2014,molesky2018,minkov2020} Simulations used a two-dimensional effective-index model at 1.55~$\mu\mathrm{m}$ under TM polarization. Each device comprised 450-nm-wide waveguides on a 1.25-$\mu\mathrm{m}$ pitch, 3.1-$\mu\mathrm{m}$ linear port tapers, and a $21\times21~\mu\mathrm{m}^2$ trainable region. The density grid was 50~nm. A 150-nm conic filter imposed the minimum length scale, and the subpixel-smoothed projection of Hammond \emph{et al.}\cite{hammond2025} with $\beta=50$ and $\eta=0.5$ drove differentiable binarization. Simulation time was 11~ps with a $10^{-4}$ shutoff threshold.

Sixteen fundamental-mode source simulations recovered the complex forward response $\mathbf{S}^{+}$ and input-side reflection $\mathbf{S}^{-}$. The electromagnetic objective was
\begin{equation}
\mathcal{L}_{\mathrm{EM}}=\left\|\mathbf{S}^{+}-\mathbf{W}_{\mathrm{target}}\right\|_{\mathrm{F}}
+\left\|\mathbf{S}^{-}\right\|_{\mathrm{F}}.
\label{eq:electromagnetic-objective}
\end{equation}
Adam used a learning rate of 0.05. Transfer fidelity was
\begin{equation}
\epsilon_{\mathrm{rel}}=\frac{\left\|\mathbf{S}^{+}-\mathbf{W}_{\mathrm{target}}\right\|_{\mathrm{F}}}
{\left\|\mathbf{W}_{\mathrm{target}}\right\|_{\mathrm{F}}}.
\label{eq:relative-matrix-error}
\end{equation}

\subsection{Phase Calibration and Robustness}

Layer-specific phase offsets $\mathbf{b}_{\ell}$ were added to each fixed simulated response,
\begin{equation}
\mathbf{a}_{\ell}=\mathbf{S}^{+}_{\ell}\left[\mathbf{a}_{\ell-1}\odot
\exp\left(i\left(\alpha\mathbf{z}+\mathbf{b}_{\ell}\right)\right)\right].
\label{eq:phase-calibration}
\end{equation}
Offsets and the training inverse-temperature $\gamma$ were calibrated on the MNIST training split for 150 full-training-set Adam updates at learning rate 0.002; the simulated matrices remained fixed and the test split did not enter these updates.

For a sample $n$ with true class $y_n$, the correct-class intensity margin was
\begin{equation}
m_n=I_{n,y_n}-\max_{c\ne y_n}I_{n,c}.
\label{eq:class-intensity-margin}
\end{equation}
Realized-layer insertion loss was $-10\log_{10}(\|\mathbf{S}^{+}_{\ell}\|_{\mathrm{F}}^2/N)$. Robustness analysis added independent complex Gaussian matrices $\mathbf{G}_{\ell,s}$,
\begin{equation}
\widetilde{\mathbf{S}}^{+}_{\ell,s}=\mathbf{S}^{+}_{\ell}+
\varepsilon\frac{\|\mathbf{S}^{+}_{\ell}\|_{\mathrm{F}}}{\|\mathbf{G}_{\ell,s}\|_{\mathrm{F}}}\mathbf{G}_{\ell,s}.
\label{eq:matrix-perturbation}
\end{equation}
Error levels ranged from 0 to 0.20 in increments of 0.025. Each nonzero level used 20 seeds; bands report the sample standard deviation.

\begin{suppinfo}
Architecture and encoding schematic; depth-dependent loss and insertion loss; optical readout and confusion matrix; CIFAR-10 readout equations; checkpoint and evaluation details (PDF).
\end{suppinfo}

\section*{Author Contributions}

Azka Maula Iskandar Muda: Conceptualization, Methodology, Software, Validation, Formal analysis, Investigation, Data curation, Visualization, Writing---original draft, Writing---review and editing. U\u{g}ur Te\u{g}in: Conceptualization, Methodology, Supervision, Project administration, Funding acquisition, Resources, Writing---review and editing.

\begin{acknowledgement}
This work was supported by a 2024 Optica Foundation Challenge Award. The authors acknowledge support from the Tidy3D Academic License Program.
\end{acknowledgement}

\section*{Data and Code Availability}

Code and workflows for regenerating the data are available at:
\par\noindent\url{https://github.com/utegin-lpt/DID-StructuralNonl}.

\section*{Conflict of Interest}

The authors declare no competing financial interest.

\end{document}